\documentclass[a4paper,11pt]{article}
\usepackage{graphicx}
\usepackage{amsmath}
\usepackage{amssymb}
\usepackage{amstext}
\large
\begin{document}
\makeatletter
\renewcommand{\@oddhead}{}
\renewcommand{\@evenhead}{}
\renewcommand{\@evenfoot}{\hbox to 16cm {Astron. Tsirkulyar
No.~1596
\hfil\thepage\hfil~~~~~~~~~~~12 March 2013}}
\renewcommand{\@oddfoot}{\hbox to 16cm {Astron. Tsirkulyar
No.~1596
\hfil\thepage\hfil~~~~~~~~~~~12 March 2013}}
\makeatother
\noindent
{ISSN 0236-2457} \hfill Proof ! \\

{\large\bf
\centerline{ASTRONOMICHESKII TSIRKULYAR}}
\medskip
\hrule
\medskip
\large
\centerline{Published by the Eurasian Astronomical Society}
\centerline{and Sternberg Astronomical Institute}
\medskip
\hrule
\medskip
\centerline{No. 1596, 
2013 March 12}
\medskip
\hrule
\bigskip
\medskip

\centerline{\bf The Solar Corona: Why It Is Interesting for Us}

\medskip

\vspace{2mm}

%
%
\noindent
{\bf Abstract.}
Strong magnetic fields are of vital importance to the physics
of the solar corona.
They easily move a rarefied coronal plasma.
Physical origin of the main structural element of the
corona, the so-called coronal streamers, is discussed.
It is shown that the reconnecting current layers inside streamers
determine their large-scale structure and evolution, including
creation, disruption and recovery.
Small-scale (fine) magnetic fields in the photosphere experience
random motion.
Their reconnection appears to be an important source of energy
flux for quiet-corona heating.
For active-corona heating, the peculiarities of entropy and
magnetoacoustic waves, related to radiative cooling, are
significant and should be taken into account in the coronal
heating theory.

\bigskip

\begin{figure}[hb]
\parbox[b]{80mm}{ 
    \caption{Composite using the White-Light total eclipse image
         of 11 July 2010 taken in French Polynesia and the
         simultaneous AIA (SDO) image of the disk put
         inside.
         The shape of the corona reveals the structure of the solar
magnetic field with open field lines at the poles of the Sun
and closed field lines above which the distribution of plasma
takes the form of {\em coronal rays\/} or
{\em coronal streamers\/} shaped like medieval helmets.
These formations are connected with the large-scale magnetic
fields on the surface of the Sun.
         Courtesy of Jean Mouette and Serge Koutchmy,
         CNRS France and AIA (SDO) from NASA}
    \noindent \label{fig1} } 
\hspace{8mm}
\parbox[b]{69mm}{ 
    \includegraphics*[height=.322\textheight]{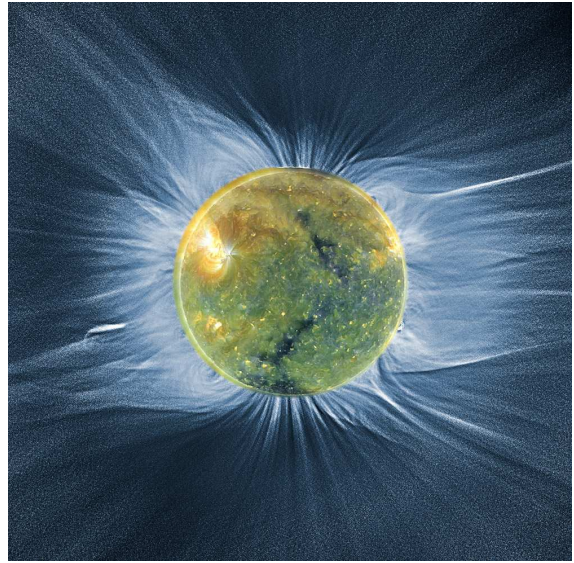}
     } 
\end{figure}

Perhaps the most amazing aspect of the corona of the Sun is its
intricate beauty.
Thousands of people move through thousands of km
(even from Moscow to Australia)
to achieve a place of the best seeing a solar eclipse.
They move heaven and earth in order to observe this
nice natural phenomenon.
Recall that the corona, consisting mainly of ionized plasma,
becomes visible to the naked eye during a total eclipse
(Fig.~\ref{fig1}).

%
%

Parker (Parker E.N., ApJ, 1958, {\bf 128}, 677)
suggested that outer parts of the corona must be expanding
in the form of a {\em solar wind\/}.
The first calculations of the magnetic field in the corona
were based on two main premises:
the magnetic field over the photosphere is potential up to a
certain height, at which the field
becomes purely radial owing to drawing-out by the solar wind.
The magnetic fields calculated under these simplified
assumptions exhibited a reasonable correlation with the optical
structure of the chromosphere and corona, as well as with the
radio- and soft X-ray pictures of the Sun.

The coronal magnetic field constructed
by this method contains neutral points where in the
presence of plasma the current layers appear
(Syrovatskii S.I., Sov. Astron. - AJ, 1962, {\bf 6}, 768).
They change geometry of magnetic field.
A current layer with quasi-radial fields of opposite
direction on either side of it appear inside a coronal
streamer similar to the current layer in the magnetosphere
tail.
In both cases the dipole magnetic field is drawn out by the
stream of the solar wind plasma:
in the corona it is the dipole magnetic field of an extended
active region, and in the magnetosphere it is the Earth magnetic
field.
The simple 2D problem of stretching out of a dipole field by a
plasma flow was formulated assuming that the field is frozen in
the flow which is accelerated similar to the solar wind
(Somov B.V. and Syrovatskii S.I., Sov. Phys. - JETP, 1972,
{\bf 34}, 992).

%
%
\begin{figure}[hb]
\parbox[b]{72mm}{ 
{The capture of the magnetic field by the solar wind occurs from the
interior of the field itself.
The plasma slowly flows along the field lines in the
strong-field region, near the dipole.
However, as the magnetic field becomes weaker
with a height in the corona, the plasma flow becomes
stronger and
is smoothly transformed into a radial solar wind that carries
an external part of the field away.
As a result, a quasi-stationary picture of magnetic field can be
established for a long-lived active region as illustrated by
Fig.~\ref{fig2}.}

\vspace{5mm}
    \caption{Magnetic field lines corresponding to the first
             analytical 2D model for coronal streamer:
    (a) the general solution with a reverse current in the
        region~RC;
    (b) the particular (stationary) solution without a reverse
        current.}
    \noindent \label{fig2} } 
\hspace{8mm}
\parbox[b]{69mm}{ 
    \includegraphics*[height=.46\textheight]{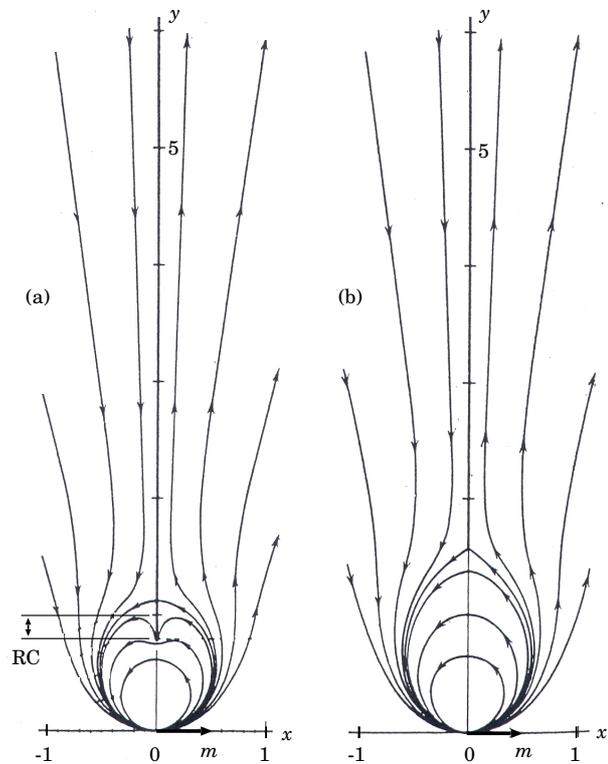}
     } 
\end{figure}

We see that the MHD approximation of a strong magnetic field can
be very good in reproducing the large-scale structures
in the corona.
Moreover, varying with time according to boundary conditions,
the magnetic field easily sets the highly conducting coronal
plasma in motion.
Its kinematics is uniquely defined by two equations.
The first of them follows from the momentum conservation law
and means that the acceleration is orthogonal to the magnetic
field lines.
The second equation is a corollary of the freezing-in condition.
For example, the set of 2D ideal MHD equations describing the
plasma flows can be rewritten as the following set of equations
(e.g., Somov B.V., Plasma Astrophysics, Part II, Reconnection and
Flares, New York, Springer SBM, 2013, Ch. 2):
\begin{equation}
    \Delta \, A = 0 \, ,
    \quad
    {d {\bf v} \over d t} \times \nabla A = 0 \, ,
    \quad
    \frac{dA}{dt} = 0 \, ,
    \quad
    {\partial \rho \over \partial t}
  + {\rm div} \, \rho {\bf v} = 0 \, .
    \label{cE}
\end{equation}
Here the scalar function $ A (x,y,t) $ is commonly called a
vector potential because of definition the vector potential
$ {\bf A}
  = \left\{ \, 0 , \, 0 , \, A \, (x, y, t) \, \right\} $
for the magnetic field $ {\bf B} = {\rm rot} \: {\bf A} $.
A complete solution of the set of equations (\ref{cE}),
including the velocity field and the plasma density distribution,
was obtained in a vicinity of a reconnecting current layer
at a hyperbolic zeroth point of magnetic field
(Somov B.V. and Syrovatskii S.I., in {\em Neutral Current Sheets
in Plasma\/}, New York and London, Consultants Bureau, 1976,
p. 13).

In the corona, more complicated models are required in order
to describe the coronal streamer behavior in the periods of
high solar activity.
A generalization of the model illustrated by
Fig.~\ref{fig2}
%
%
is needed because the current layer inside a
streamer can be disrupted into parallel current filaments or
ribbons
(Wagner S.A. and Somov B.V., in {\em Cosmicheskie Issledovania\/},
Sankt-Peterburg, FTI, 1991, p. 79, in Russian).

\begin{figure}[hb]
\parbox[b]{70mm}{ 
    \caption{Magnetic field lines corresponding to the
             generalized 2D model for a coronal streamer with
             non-equilibrium RCL.
             The effective magnetic `charges' $ {\rm e}_{\rm n} $
             and $ {\rm e}_{\rm s} $ at the points $ x = \pm \, a $
             model the photospheric (or
             under-photospheric) sources of magnetic field.
    (a) Disruption of the RCL due to magnetic reconnection at
        the point $ h_{ \rm X } $.
        At the point $ h_{ \rm Y } $ the magnetic force equals
        zero, but at the points $ h_{ \rm D } $ and $ h_{ \rm U } $
        it is not equal to zero and is directed downwards and
        upwards respectively.
        Therefore, fast reconnection is driven by the
        magnetic forces $ {\bf F}_{\rm mag} $, acting on the edges
        of the gap.
    (b) The non-stationary process of recovery of the RCL
        via a secondary reconnecting current layer
        ($ h_{ \rm Y1 } , h_{ \rm Y2 } $).
        Thick empty arrows show the plasma flows in the
        vicinity of this new RCL.
        $ {\bf V}_{\rm rec} $ is the velocity corresponding to
        the reconnection rate in the secondary RCL.
        }
    \noindent \label{fig3} } 
\hspace{4mm}
\parbox[b]{69mm}{ 
    \includegraphics*[height=.48\textheight]{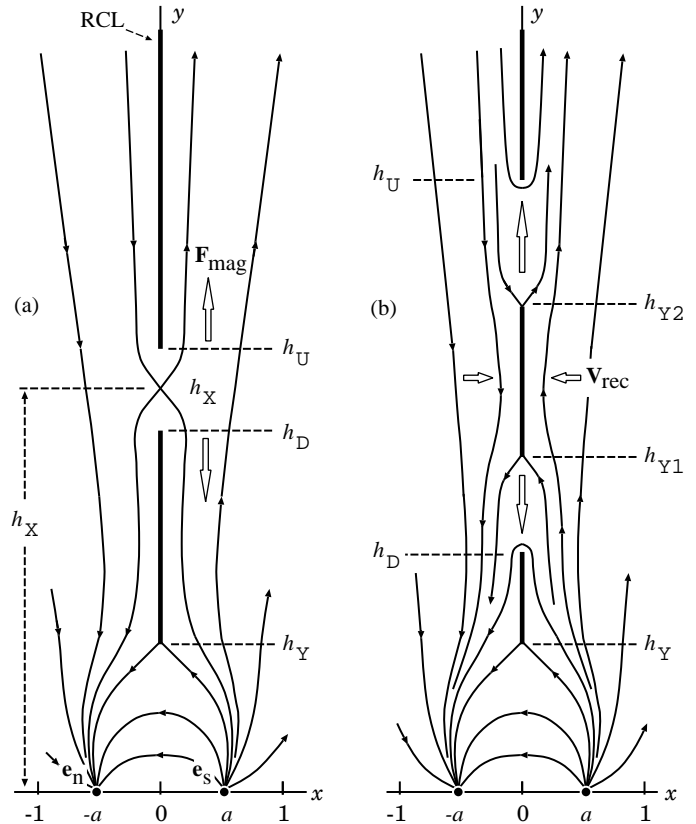}
     } 
\end{figure}

Fig.~\ref{fig3}a
%
%
demonstrates such a model which assumes that a rupture (a gap
between points $ h_{ \rm D } $ and $ h_{ \rm U } $)
of the reconnecting current layer (RCL, two thick vertical
segments) emerges in a region of high electric resistivity,
for example, anomalous resistivity
due to the excitation of plasma turbulence.
Fast magnetic reconnection takes place in the vicinity of
the X-type zeroth point $ h_{ \rm X } $ of a strong magnetic
field.
The reconnection process is driven by the uncompensated
magnetic forces $ {\bf F}_{\rm mag} $, acting on the edges
of the gap, $ h_{ \rm D } $ and $ h_{ \rm U } $,
and having a disruptive influence on the RCL.
A simple analytical model of a disrupting RCL
(Somov B.V. and Syrovatskii S.I., Bull. Acad. Sci. USSR, Phys.
Ser., 1975, {\bf 39}, No.~2, 109)
shows that the magnetic tension forces $ {\bf F}_{\rm mag} $
are proportional the the size of the gap,
$ h_{ \rm U } - h_{ \rm D } $,
and are tending to increase it.

%
%

While considering the $ ( x, y ) $~plane as a complex
plane~$ z = x+{\rm i} y $, we relate an
analytic function~$ F $ to the vector potential~$ A $
as follows
$ F ( z, t )
  = A ( x, y, t) + {\rm i} \, A^{+} ( x, y, t ) \, . $
Then
\begin{displaymath}
    {\bf B} = B_{x} + {\rm i} \, B_{y} = - \, {\rm i}
    \left( { dF } / { dz } \right)^{*} ,
    \label{crB1}
\end{displaymath}
where the asterisk denotes the complex conjugation.
Define
$ {\bf B}^{*} = B_{x} - {\rm i} \, B_{y} \equiv B (z) $.
The magnetic field of non-equilibrium disruptive streamer shown in
Fig. \ref{fig3}a
%
%
is given by formula
\begin{equation}
    B (z) = c_1
    ( a^2 - z^2 )^{-1}
    ( z^2 + h_{ \rm X }^2 )
    ( z^2 + h_{ \rm Y }^2 )^{1/2}
    ( z^2 + h_{ \rm D }^2 )^{-1/2}
    ( z^2 + h_{ \rm U }^2 )^{-1/2} .
    \label{cE-2}
\end{equation}

The X-type point $ h_{ \rm X } $ at the center of the reconnection
region has a special status.
If the plasma density near this point does not drop too much in
the reconnection process (see, however,
Somov B.V. and Syrovatskii S.I., in {\em Neutral Current Sheets
in Plasma\/}, New York and London, Consultants Bureau, 1976,
Ch. 3, Sect. 2),
then a secondary current layer (the thick vertical segment between
points $ h_{ \rm Y1 } $ and $ h_{ \rm Y2 } $ in
Fig.~\ref{fig3}b)
%
%
will appear.
Otherwise, the plasma is not enough to produce the secondary
current layer capable of suppressing the current layer disruption.
In other words, the primary reconnecting current layer (RCL) can
be completely disrupted or, alternatively, recreated in the
non-stationary process shown in
Fig.~\ref{fig3}b.
%
%
The streamer will make a full recovery from the rupture
(Fig.~\ref{fig3}a)
%
%
to its original shape
(Fig.~\ref{fig-2-0311}).
%
%
%
The magnetic field of a recovering streamer
(Fig.~\ref{fig3}b)
%
%
is
\begin{equation}
    B (z) = c_2
    ( a^2 - z^2 )^{-1}
    ( z^2 + h_{ \rm Y }^2 )^{1/2}
    ( z^2 + h_{ \rm Y1 }^2 )^{1/2}
    ( z^2 + h_{ \rm Y2 }^2 )^{1/2}
    ( z^2 + h_{ \rm D }^2 )^{-1/2}
    ( z^2 + h_{ \rm U }^2 )^{-1/2} .
    \label{cE-2}
\end{equation}

So, the basic idea articulated above is that coronal streamer
formation is a twofold magnetic process.
First, the magnetic field plays a passive role in shaping
streamers by some processes involving the stretching-out the field
by the solar wind acceleration and motion.
Second, the magnetic field plays an active role in providing
dynamic behavior of a streamer by magnetic reconnection.
The non-stationary dynamics of a coronal streamer combines two
opposite processes:
(a) the disruption of a reconnecting current layer
inside a streamer and
(b) its recreation, which we call recovery.

As a consequence, depending on physical conditions, a streamer
can be completely disrupted and disappears or be recovered once or
several times after being disrupted.
That is why a streamer can exist a very long time even as long as
an underlying active region exists.
Moreover, its large-scale external structure looks like the same
stationary configuration.
Non-stationary plasma flows inside a streamer related to magnetic
reconnection in a recovering streamer
(Fig.~\ref{fig3}b)
%
%
are always present but not always they are well observable.
Because of conservation of the global configuration, a coronal
streamer may be compared with a river: one cannot come in the
same river twice.

The orientation of coronal streamers, the change of shape of the
corona with the solar activity cycle, etc.,
all these observational facts tell us about the existence of
coronal magnetic fields.
However it is quite difficult to measure them because the coronal
emission is exceedingly faint.
Until the present, observations are scarce and our knowledge about
the coronal magnetic field comes mainly from the theoretical
extrapolation of photospheric fields and from comparison of the
theoretical model predictions with the observed large-scale
structures of the corona.
\begin{displaymath}
    \ast \quad  \,\, \ast \quad \,\, \ast
\end{displaymath}

Fine structure of solar magnetic fields presumably has properties
of complex field configurations containing many
places (points or lines) where reconnection occurs.
Such a situation frequently appears in astrophysical plasmas,
for example in a set of closely packed flux tubes suggested by
Parker~(Parker E.N., ApJ, 1972, {\bf 174}, 499).
The tubes tend to form many reconnecting current layers (RCLs)
at their interfaces.
This may be the case of active regions when the field-line
foot\-point motions are slow enough to consider the evolution of
the coronal magnetic field as a series of equilibria, but fast
enough to explain {\em coronal heating\/}.

Magnetic flux tubes in the photosphere are subject to constant
buffeting by convective motions, and as a result, flux tubes
experience random walk through the photosphere.
From time to time, these motions will have the effect that a flux
tube will come into contact with another tube of opposite polarity.
We refer to this process as reconnection in weakly-ionized
plasma
(Litvinenko Yu.E. and Somov B.V., Solar Phys., 1994, {\bf 151},
265).
Another possibility is the photospheric dynamo effect
(H\'enoux J.C. and Somov B.V., A\&A, 1997, {\bf 318}, 947)
which, in an initially weak field, generates thin flux tubes of
strong magnetic fields.
Such tubes extend high into the chromosphere and contribute to
the mass and energy balance of the quiet corona.

{\em SOHO\/}'s MDI
observations have shown that the magnetic field in the quiet
network of the photosphere is organized into relatively small
`concentrations' (magnetic elements, small loops etc.)
with fluxes in the range of $ 10^{18} $~Mx up to a few times
$ 10^{19} $~Mx, and an intrinsic field strength of the order of
a kilo\-gauss.
These concentrations are embedded in a super\-posion of flows,
including the granulation and super\-granulation.
They {\em fragment\/} in response to sheared flows,
{\em merge\/} when they collide with others of the same
polarity, or {\em cancel\/} against concentrations of opposite
polarity.
Newly emerging fluxes replace the canceled ones.

Direct evidence that the so-called `{\em magnetic carpet\/}'
(Day C., Physics Today,~1998, March issue, 19),
an ensemble of magnetic concentrations in the photosphere,
really can heat the corona comes from the two other
{\em SOHO\/} instruments: CDS and EIT.
Both have recorded local brightenings of hot plasma
that coincide with disappearances of the carpet's elements.
This indicates that just about all the elements reconnect and
cancel, thereby releasing magnetic energy.

The transition region and chromospheric lines observed
by {\em SOHO\/} together with radio emission of the
quiet Sun simultaneously observed by VLA show that
the corona above the magnetic network has a higher
pressure and is more variable than that above the interior of
super\-granular cells.
Comparison of multi\-wave\-length observations of quiet
Sun emission shows good spatial correlations between enhanced
radiations originating from the chromosphere to the corona.
Furthermore
the coronal heating events follow the basic properties of
regular solar flares and thus may be well interpreted as
{\em micro\-flares\/} and {\em nano\-flares\/}.
The differences is mainly quantitative
(Krucker S. and Benz A.O., Solar Phys.,~2000, {\bf 191}, 341).

What do we really need to replenish the entire magnetic carpet
quickly, say 1-3~days? --
A rapid replenishment, including the entire cancelation of
magnetic fluxes, requires the fundamental assumption of a
{\em two-level\/} reconnection in the solar atmosphere
(Somov B.V., Bull. Russ. Acad. Sci.,~1999, {\bf 63}, 1157).
First,
we apply the concept of fast reconnection of electric currents
as the source of energy for micro\-flares to explain coronal
heating in quiet regions
(Somov B.V. and H\'enoux J.-C., in
{\em Magnetic Fields and Solar Processes\/},
9th Eur. Meet. on Solar Phys., ESA SP-448, 1999, 659).
Second,
in addition to coronal reconnection, we need an efficient
mechanism of magnetic field and current dissipation in the
photosphere.
The presence of a huge amount of neutrals in the weakly ionized
plasma in the temperature minimum region makes its
properties very different from an ideal MHD medium.
Dissipative {\em collisional\/} reconnection is very efficient here
(Litvinenko and Somov,~1994).
\begin{displaymath}
    \ast \quad  \,\, \ast \quad \,\, \ast
\end{displaymath}

While the corona is evidently heated everywhere, there is no
question that it is heated most intensively within active
regions where the magnetic field is the strongest.
Detailed models of coronal heating in active regions typically
invoke mechanisms belonging to one of the
two broadly defined categories: wave (AC) or stress (DC)
heating.
In the AC heating,
the large-scale magnetic field serves essentially as a conductor
for small-scale MHD waves propagating into the corona.
Thus the properties of these waves are of principal importance.

In the corona, the low-frequency MHD oscillations can be studied
comprehensively.
They are observed almost at all wavelengths (see
Asch\-wan\-den M.J., Physics of the Solar Corona, Berlin,
Springer,2004).
Most of these oscillations are commonly interpreted as standing
oscillations of various types in coronal magnetic loops.
Meanwhile the oscillations of coronal loops observed from
{\em TRACE\/} satellite in EUV are, as a rule, damped rapidly.
The ratio of the characteristic damping time~$ \tau_{d} $ to the
oscillation period~$ \tau_{\omega} $ is
$ \tau_{d} / \tau_{\omega} = 1.8 \pm 0.8 $ in the range of
periods $ \tau_{\omega} = 317 \pm 114 $~s.
Such {\em rapid damping\/} of the MHD oscillations seemed difficult
to explain.

Why rapidly damped oscillations are seen best in a small group
of loops precisely in EUV radiation is a key question.
Contrary to popular belief, the answer is simple.
Where the rate of energy losses via optically thin plasma
radiation has a maximum (i.e. at $ T \sim 10^5 $~K), the
brightness of the oscillating loops also has a maximum
(i.e. in EUV) and, as a consequence, the MHD oscillations are
damped more rapidly than in other places.
This is the case of {\em slow\/} magneto\-acoustic waves
(Somov B.V., Dzhalilov N.S., and Staude J., Astron. Lett., 2007,
{\bf 33}, 309).
The significant advantage of slow magneto\-acoustic waves over
{\em fast\/} ones is that the regions of reduced magnetic field
in the former coincide with the regions of enhanced plasma
density.
Here the rapid radiative losses manifest themselves.
Meanwhile, as calculations show, fast magneto\-acoustic
waves radiate little and, therefore, are damped too slowly.

Another feature of small MHD perturbations in an optically thin,
perfectly conducting plasma with a cosmic abundance of elements is
an instability of entropy waves.
The instability mechanism is simple.
In the temperature regions of a rapid decrease in the radiative
loss function with temperature, a small decrease in temperature
causes a large increase in the rate of radiative energy losses.
Conversely a small increase in temperature is accompanied by a
decrease in the rate of radiative plasma cooling.
As a result, small perturbations grow rapidly.
The growth time for the entropy waves in the corona can vary
over a wide range: from tenths of a second to tens minutes.

The fact that the instability condition for entropy waves is almost
independent of the magnetic field strength and configuration is
fundamentally important for the theory of coronal heating.
This means that among the various physical processes involved in
the coronal heating, the growth of entropy waves can manifest itself
everywhere.
The peculiarities of entropy and magneto\-acoustic waves,
related to radiative losses of energy,  should be taken into
account in general theory of evolu\-tionarity of MHD discontinuities
(see Ch.~17 in Somov B.V., Plasma Astrophysics, Part I,
Fundamentals and Practice, New York, Springer SBM, 2013).

\vspace{2mm}

\medskip
\noindent
P.K.\,Sternberg Astronomical Institute  \hspace{8cm} \\
M.V.\,Lomonosov Moscow State University \\
13 Universitetskij prosp.\hspace{8.6cm} Boris V. Somov \\
Moscow, 119991 Russian Federation \\
{\it somov@sai.msu.ru}  \\
Received March 12, 2013

\end{document}